\documentclass[a4paper,fleqn,usenatbib,usedcolumn]{mnras}

\usepackage{newtxtext,newtxmath}

\usepackage[T1]{fontenc}
\usepackage{ae,aecompl}
\usepackage{xcolor}

\usepackage{graphicx}	
\usepackage{amsmath}	
\usepackage{multirow}
\usepackage{cases}
\usepackage{float}

\hypersetup{pdfauthor={W. N. Alston},
               pdftitle={Another variability paper},
               pdfkeywords={accretion, accretion discs -- X-rays: binaries -- X-rays: galaxies -- ULXs},
               bookmarksnumbered=true}

   \volume{{\rm in press}}
   
\def\apj{ApJ}
\def\mnras{MNRAS}
\def\apjl{ApJ}

\def\aap{A\&A}

\def\apjs{ApJS}
\def\nat{Nature}

\def\araa{Ann. Rev. A\&A}
\def\xmm{{\it XMM-Newton~\/}}
\def\xmmns{{\it XMM-Newton}}
\newcommand{\xmmn}{{\it XMM-Newton~\/}}

\def\Ledd{\hbox{$L_{\rm Edd}$}}

\def\ergs{\hbox{${\rm erg}~{\rm s}^{-1}$}}

\newcommand{\ngc}{NGC 247 ULX-1~\/}

\defcitealias{alston19a}{A19}
\defcitealias{uttleymchardyvaughan05}{UMV05}
\defcitealias{uttleymchardyvaughan17}{UMV17}

\def\gsim{\mathrel{\hbox{\rlap{\hbox{\lower4pt\hbox{$\sim$}}}\hbox{$>$}}}}
\def\lsim{\mathrel{\hbox{\rlap{\hbox{\lower4pt\hbox{$\sim$}}}\hbox{$<$}}}}

\title[Quasi-periodic dipping in NGC 247 ULX-1]{Quasi-periodic dipping in the ultraluminous X-ray source,\\ NGC 247 ULX-1}

\author[W. N. Alston]{W. N. Alston$^{1}$\thanks{E-mail: walston@sciops.esa.int}, C. Pinto$^{2}$, D. Barret$^{3}$, A. D'A\`{i}$^{2}$, M. Del Santo$^{2}$, H. Earnshaw$^{4}$, 
\and
A. C. Fabian$^{5}$, F. Fuerst$^{1}$, E. Kara$^{6}$, P. Kosec$^{6}$, M. J. Middleton$^{7}$, M. L. Parker$^{5}$,  
\and
F. Pintore$^{8}$, A. Robba$^{2,9}$, T. P. Roberts$^{10}$, R. Sathyaprakash$^{11}$, D. Walton$^{5}$, E. Ambrosi$^{2}$\\
$^{1}$European Space Agency (ESA), European Space Astronomy Centre (ESAC), Villanueva de la Ca\~{n}ada, Madrid, E-28691, Spain\\
$^{2}$INAF/IASF Palermo, via Ugo La Malfa 153, I-90146 Palermo, Italy\\
$^{3}$Institut de Recherche en Astrophysique et Planétologie (IRAP), 9 Avenue du Colonel Roche, BP 44346, 31028, Toulouse Cedex 4, France\\
$^{4}$Cahill Center for Astronomy and Astrophysics, California Institute of Technology, Pasadena, CA 91125, USA\\
$^{5}$Institute of Astronomy, Madingley Rd, Cambridge, CB3 0HA.\\
$^{6}$MIT Kavli Institute for Astrophysics and Space Research, Cambridge, MA 02139, USA\\ 
$^{7}$Department of Physics and Astronomy, University of Southampton, Highfield, Southampton SO17 1BJ, UK\\
$^{8}$INAF-IASF Milano, via A. Corti 12, I-20133 Milano, Italy\\
$^{9}$Università degli Studi di Palermo, Dipartimento di Fisica e Chimica, via Archirafi 36, I-90123 Palermo, Italy\\
$^{10}$Centre for Extragalactic Astronomy, Department of Physics, Durham University, South Road, Durham, DH1 3LE\\
$^{11}$Institut de Ciències de l`Espai, Carrer de Can Magrans, 08193, Cerdanyola del Vallès, Barcelona\\
}

\date{Accepted 2021 May 18. Received 2021 May 18; in original form 2021 April 20.}

\pubyear{2021}

\begin{document}
\label{firstpage}
\pagerange{\pageref{firstpage}--\pageref{lastpage}}
\maketitle

\begin{abstract}
Most ultraluminous X-ray sources (ULXs) are believed to be stellar mass black holes or neutron stars accreting beyond the Eddington limit.  Determining the nature of the compact object and the accretion mode from broadband spectroscopy is currently a challenge, but the observed timing properties provide insight into the compact object and details of the geometry and accretion processes.  Here we report a timing analysis for an $800$\,ks \xmmn campaign on the supersoft ultraluminous X-ray source, NGC 247 ULX-1.  Deep and frequent dips occur in the X-ray light curve, with the amplitude increasing with increasing energy band.  Power spectra and coherence analysis reveals the dipping preferentially occurs on $\sim 5$\,ks and $\sim 10$\,ks timescales.  The dips can be caused by either the occultation of the central X-ray source by an optically thick structure, such as warping of the accretion disc, or from obscuration by a wind launched from the accretion disc, or both.  This behaviour supports the idea that supersoft ULXs are viewed close to edge-on to the accretion disc.
\end{abstract}





\section{Introduction}


Ultraluminous X-ray sources (ULXs) are X-ray bright, off-nuclear sources with luminosity $\gsim 10^{39}$\,\ergs (see \citealt{Kaaret2017} for a review).  Proposed explanations for these luminous sources include i. sub-Eddington accreting stellar mass compact objects whose radiation is beamed into our line of sight (\citealt{King2001}); ii. stellar-mass BHs in a super-Eddington regime (\citealt{Poutanen2007}) or, iii. intermediate-mass black holes accreting at sub-Eddington rates from low-mass companion stars (\citealt{colbertmushotzky1999}, \citealt{Farrell2009}).

\begin{figure*}
\centering
\hspace*{-10pt}
\includegraphics[width=1.02\textwidth,angle=0]{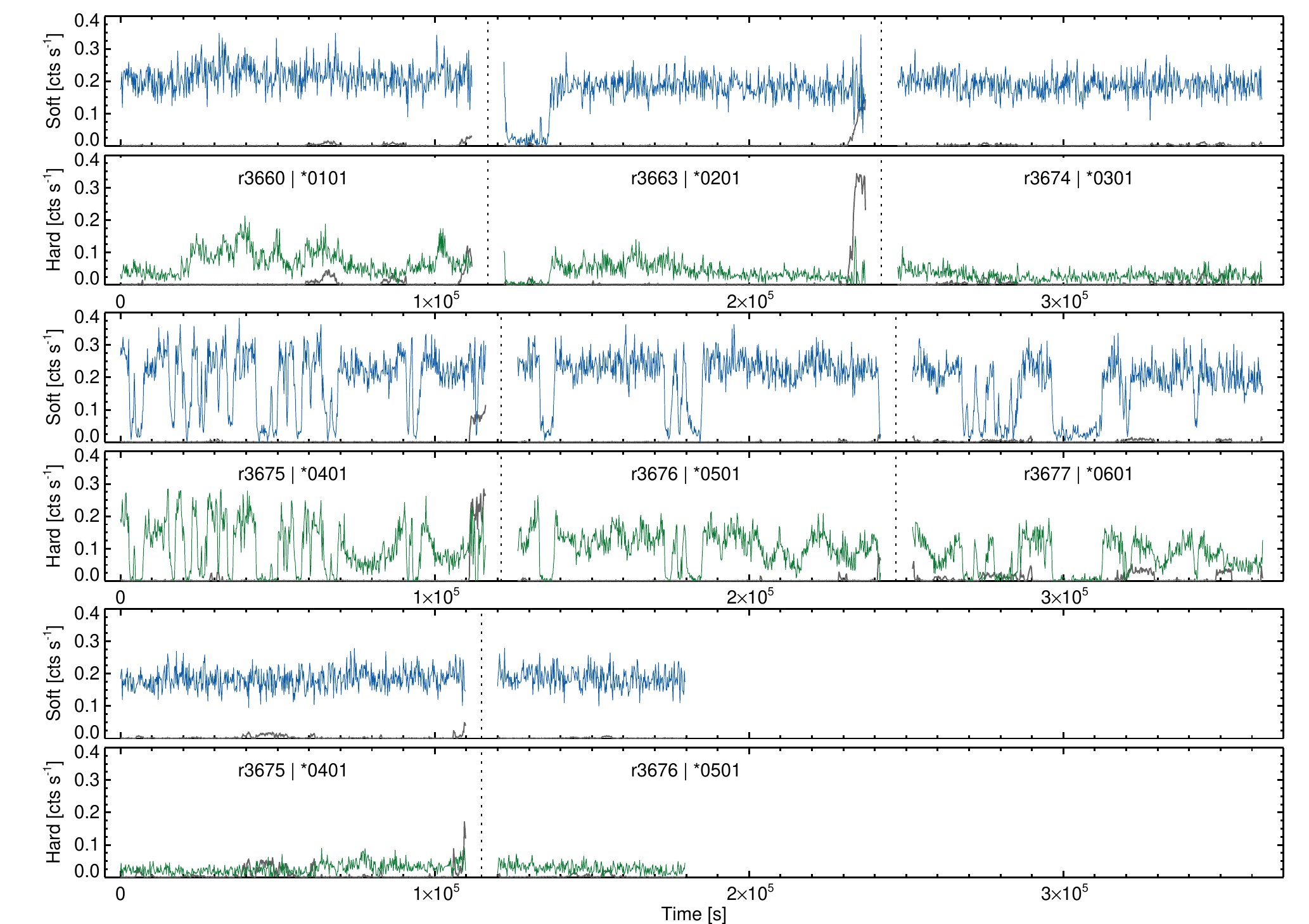}
\vspace*{-0pt}
\caption{Background subtracted light curves for 8 new \xmmn observations with $dt = 200$\,s.  The soft ($0.3-1.0$\,keV) and hard ($1.0-4.0$\,keV) bands are shown in blue and green respectively.  The grey curve shows the estimated background contribution.  Observations are separated by a vertical dotted line for plotting purposes, and are denoted by their \xmmn revolution number (e.g. r$3660$) and {\tt OBSID} (where $* = 084486$).}
\label{fig:lc}
\end{figure*}

The recent detection of coherent pulsations in several ULXs indicates they are powered by neutron stars (NS) accreting at very high Eddington rates, with luminosities up to $\sim 500 \Ledd$ (\citealt{Bachetti2014}, \citealt{Israel2017a,Israel2017b}, \citealt{Carpano2018}; \citealt{Sathyaprakash2019a}, \citealt{RodriguezCastillo2019}, \citealt{Chandra2020pulx}).  Pulsations are not always detectable as high count rates and pulsed fraction are required.  In the absence of pulsations, spectral analysis alone makes it difficult to distinguish between a BH or NS candidate (e.g. NGC 1313 X-1, \citealt{walton2020ngc1313}), which means that NSs are likely numerous among the compact objects of ULXs (\citealt{KingLasota2019}; \citealt{Pintore2017}; \citealt{Walton2018a}; \citealt{middletonking2017}; \citealt{Wiktorowicz2019ulxpop}).  

Fourier timing techniques are powerful tools for investigating in a model-independent way the geometry of accretion discs, the size scales involved, and the nature of the compact objects (e.g. \citealt{middleton07, heil09, Pinto2017, Kara2020}).  For instance, the presence of time lags between certain energy bands identifies different regions of X-ray emission in the accretion disc. The soft time lags found in three ULXs indicates that substantial Compton down-scattering is occurring in the outer disk (e.g., \citealt{demarco13}; \citealt{Pinto2017}; \citealt{Kara2020}).  Power spectral density (PSD) analysis allows us to distinguish different accretion processes occurring at various accretion rate and identify different zones within the accretion disc.


Dipping behaviour in the X-ray light curve have also been observed in some ULXs, e.g. NGC 55 (\citealt{stobbart2004}), M 94 (\citealt{lin2013dips}), NGC 628 (\citealt{liu2005dips}), NGC 5408 X-1 (\citealt{pashamstrohmayer2013, grise2013dips} and M 51 ULX-7 (\citealt{hu2021dips, Vasilopoulos2021dips}).  The number of dips observed in these sources has so far been limited, however they suggest we are viewing these sources at high inclination or could be evidence for a precessing disc.

NGC 247 ULX-1 is a supersoft ultraluminous X-ray source.  It is unique in that it switches between the soft ultraluminous X-ray state (SUL) and the supersoft ultraluminous state (ULS or SSUL, e.g. \citealt{Feng2016}).  Two previous short ($\sim 30$\,ks), \xmmn observations revealed strong dips in source flux in the light curve.  These dips lasted several ks with the flux decreasing by an order of magnitude (\citealt{Feng2016}).  In this paper we report a timing study of the dipping in the light curves of \ngc using a recently awarded $800$\,ks \xmmn campaign (PI: Pinto).  PSD analysis reveals multiple preferred periodic-like timescales for the dips.  A detailed spectral analysis is provided in two companion papers: \citealt{pinto2021_ngc247sub} on the high-resolution spectroscopy and D'A\`{i} et al. \textit{in prep} on the broadband spectroscopy.


\section{Observations and data analysis}
\label{sec:obs}

NGC 247 ULX-1 was observed 8 times by \xmm over 1 month, starting 2019-12-03 (see \citealt{pinto2021_ngc247sub} for observation details).  Seven observations have duration $\sim 115$\,ks and one lasts for $\sim 55$\,ks.  Two short ($\sim 35$\,ks) observations from 2016 are not included in this analysis as they are too short to probe Fourier frequencies below $\sim 1 \times 10^{-4}$\,Hz.  

The EPIC-pn Observation Data Files (ODFs) were processed following standard procedures using the \xmm\ Science Analysis System (SAS; v19.0.0), using the conditions {\tt PATTERN} 0--4 and {\tt FLAG} = 0.  Source counts were extracted from a circular region with radius 20 arc seconds.  The background is extracted from a large rectangular region on the same chip, avoiding the Cu ring on the outer parts of the chip.  Following \citet{alston13b,alston19a} for flares with duration $\le 200$\,s, the source light curve was cut out and interpolated between the gap by adding Poisson noise using the mean of neighbouring points.  The interpolation fraction was typically $< 0.3$\,\%.  High background segments, typically towards the end of the revolutions (e.g. revolutions (revs)~$3633,3675$), were excluded from the analysis.

Figure~\ref{fig:lc} shows the background subtracted light curves for the $0.3-1.0$\,keV (hereafter soft) and $1.0-4.0$\,keV (hereafter hard) bands.  These bands are motivated by broadband spectral analysis in \citealt{pinto2021_ngc247sub} and D'Ai et al. \textit{in prep}.  Above $4.0$\,keV the light curve is dominated by the background so we exclude it from the PSD analysis in Sec.~\ref{sec:psd}, however these bands are used where appropriate in Sec.~\ref{sec:varspec}. 

Clear dipping behaviour is apparent in four of the observations (revs~$3663,3675,3676,3677$), with the dip duration ranging from $~\sim 1-20$\,ks.  Four observations do not show any obvious dipping behaviour: revs~$3660,3674,3678,3680$.  We therefore divide the observations into two data sets - the `dip' and `non-dip' segments.  With the exception of rev~$3663$, the dip segments have higher source flux and are spectrally harder (see also D'Ai et al. \textit{in prep}, \citealt{pinto2021_ngc247sub}).  In the following analysis, we use a segment length of $110$\,ks for the dip epochs and $55$\,ks for the no-dip epochs, owing to the shorter duration of rev~$3680$.


\section{Timing analysis}

\subsection{The power spectrum}
\label{sec:psd}

    \begin{figure}
            \hspace*{-1pt}
            \includegraphics[width=0.44\textwidth,angle=0]{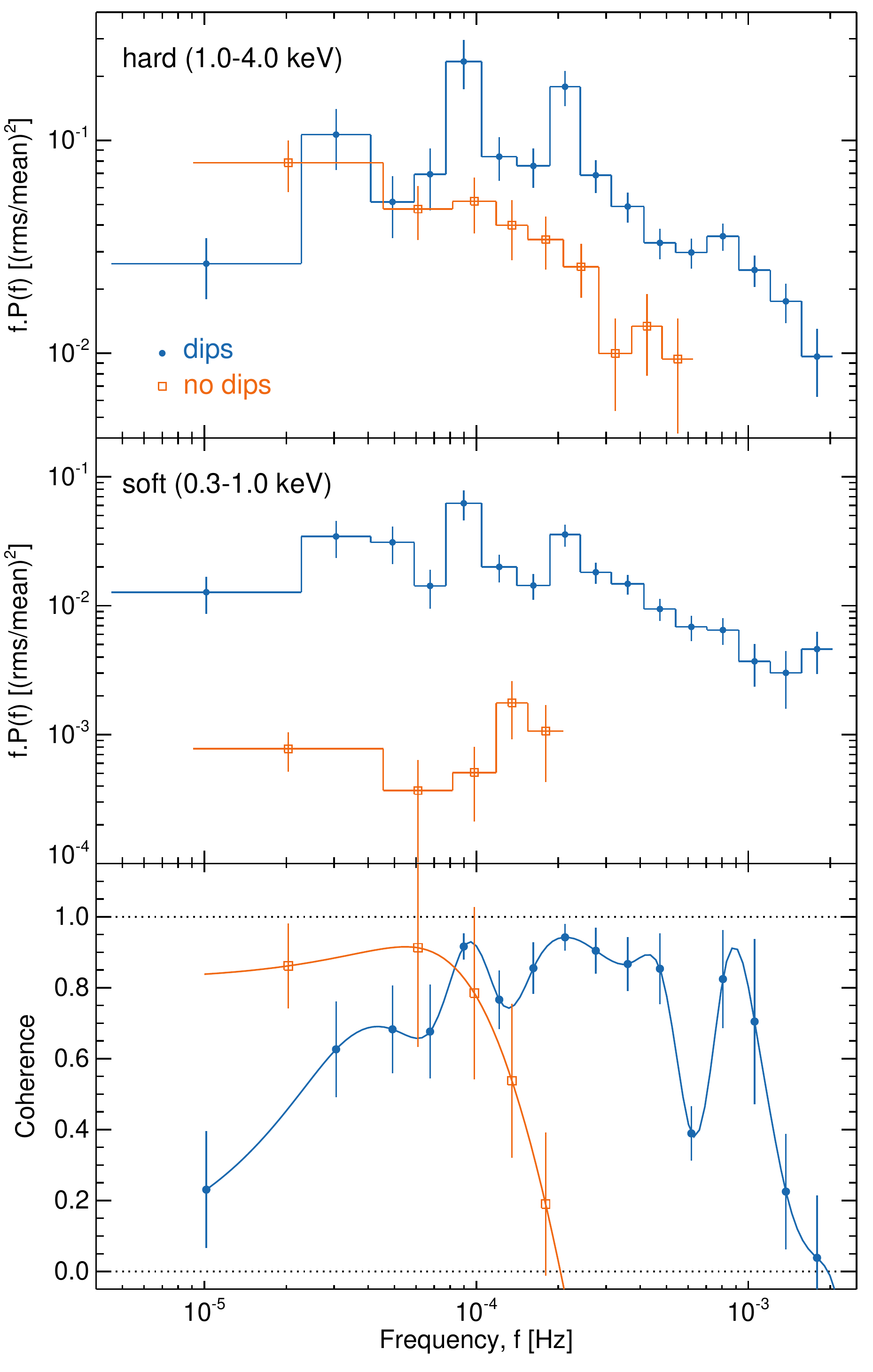}
            \vspace*{-4pt}
        \caption{The Poison-noise subtracted PSDs in $\nu P(\nu)$ units for the `dips' (blue circles) and `no dips' (orange open squares) segments.  The upper panel shows the hard band ($1.0-4.0$\,keV) and the middle panel shows the soft band ($0.3-1.0$\,keV). The lower panel shows the coherence between the hard and soft bands. A simple cubic spline has been overlaid to highlight the frequency dependence.}
        \label{fig:psdcoh}
    \end{figure}

The PSD was estimated using the standard method of calculating the periodogram (e.g. \citealt{priestley81}), with an ${\rm [rms/mean]}^2$ normalisation, then averaging over $M$ segments at each Fourier frequency.  Non-overlapping neighbouring frequency bins are then averaged (see e.g. \citealt{vaughan03a}).  In order to pick out features over a broad range of frequencies we used a geometrically averaged frequency bins, each bin spanning a factor $\gsim 1.3$ in frequency larger than the previous bin.  Figure~\ref{fig:psdcoh} shows the soft and hard band noise-subtracted PSDs in units of $\nu P(\nu)$ for both the dipping and non-dipping segments (the noise is estimated from standard formulae, e.g. \citealt{vaughan03a}).

The dipping observations show clear peaks in power at $\sim 1 \times 10^{-4}$\,Hz and $\sim 2 \times 10^{-4}$\,Hz.  A further broader peak can be seen at $\sim 8 \times 10^{-4}$\,Hz.  At frequencies below $\sim 2 \times 10^{-5}$\,Hz the PSD is rolling over towards zero power.  With the exception of the appearance of narrow peaks in power, the shape of the dip and no-dipping segments in the hard band looks similar.   The non-dip PSD has a smooth power-law like frequency dependence, with a steeper slope in the hard band.

\subsection{Coherent variations}

We study the correlated variations between the two bands using the Fourier coherence.  This provides a measure of the linear correlation between two time series $x(t), y(t)$ as a function of Fourier frequency, in the range [0,1] (\citealt{vaughannowak97}).  We estimated cross-spectrum, $C_{xy}(f)$, by first averaging the complex values over $M$ non-overlapping segments, then averaging over neighbouring frequency bins, in a similar manner to the PSD.  This amplitude part of the cross-spectrum gives the coherence, $\gamma^2$, which is then corrected for Poisson noise, see e.g. \citealt{uttley14rev} for more details.

The coherence is shown in Figure~\ref{fig:psdcoh} for both the dipping and non-dipping light curves.  A simple cubic spline has been fit to the data to illustrate the frequency dependence of the data. The coherence is high ($\gsim 0.9$) at frequencies where the soft and hard band PSDs peak, and drops at frequencies between these peaks in power.  In contrast, a high coherence at low frequencies is observed in the non-dipping light curves, which falls off sharply to zero at $\sim 2 \times 10^{-4}$\,Hz, most likely caused by the onset of Poisson noise at high frequencies.

The phase part of the cross-spectrum provides an estimate of the frequency dependent delay between $x(t), y(t)$, which can be converted to a time delay in seconds (\citealt{vaughannowak97}).  We estimated this for both the dipping and non-dipping segments and found the delay is consistent with zero lag at all Fourier frequencies.

\subsection{Modelling the power spectra}
\label{sec:psdmod}

\begin{figure}
\centering
    \vspace*{-4pt}
    \hspace*{-8pt}
    \includegraphics[width=0.46\textwidth,angle=0]{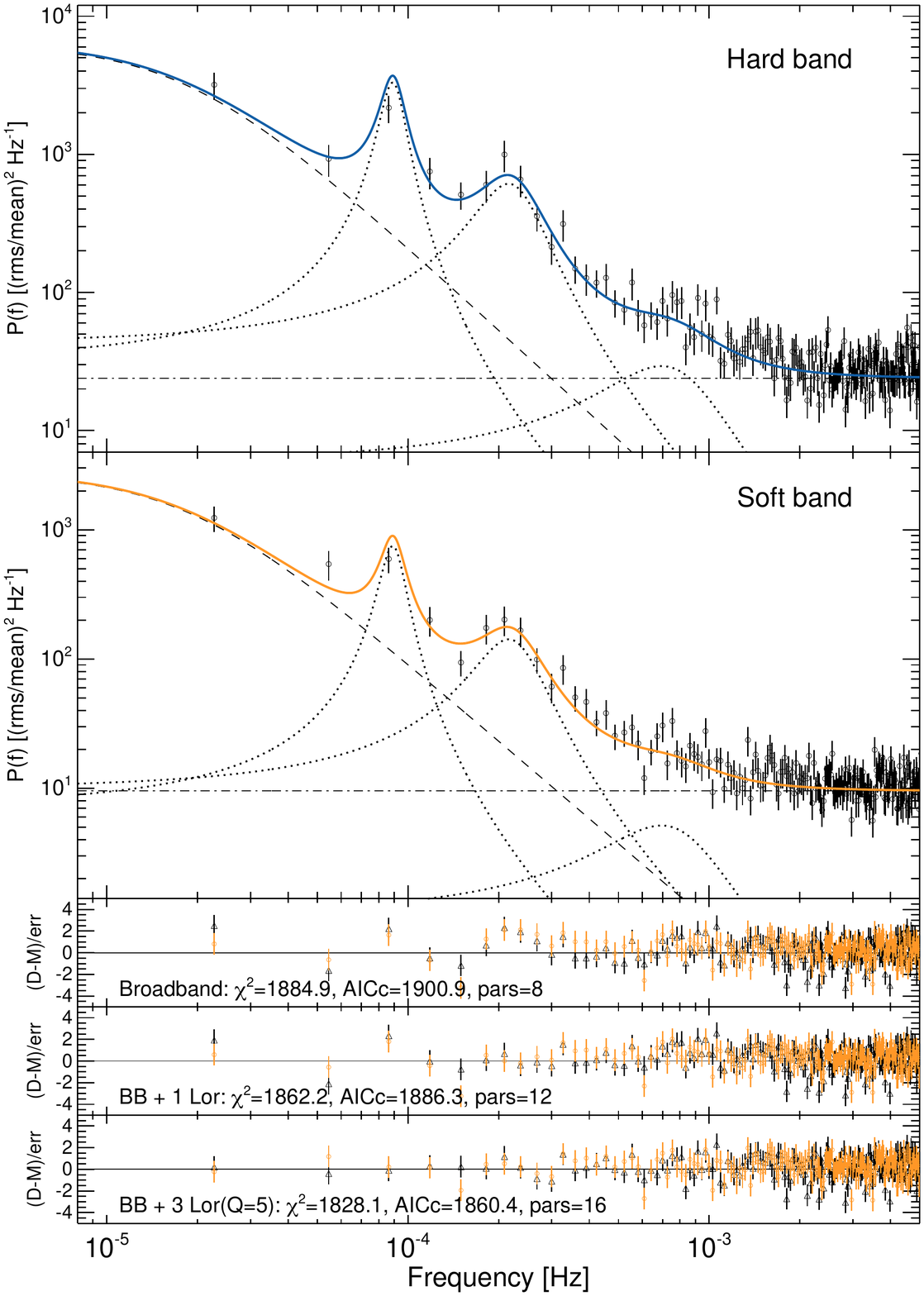}
    \vspace*{-4pt}
    \caption{PSD modelling of the dipping observations using multiple Lorentzian components.  The best fit model (BB+3Lor($Q_{\rm l_1}=5$) with a zero-centered Lorentzian (dashed) and three additional narrower Lorentzians (dotted) are shown for the hard and soft bands in the upper and middle panels respectively.  The horizontal dot-dashed line is the Poisson noise parameter.  The lower three panels show the standard residuals for this and two less complex models.}
    \label{fig:psdmod}
\end{figure}

To characterise the observed peaks in variability power in the dip segments we modelled the PSD using multiple Lorentzian functions;

\begin{equation}
\label{eqn:lorentz}
   P(\nu) = \frac{N_{\rm l} (\sigma_{\rm l} / 2 \pi) }{(\nu - {\nu}_{\rm c})^2 + (\sigma_{\rm l} / 2)^2}
\end{equation}

\noindent where ${\nu}_{\rm c}$ and $\sigma_{\rm l}$ are the line centroid and width parameters, respectively, and $N_{\rm l}$ is a normalisation.  The quantity $Q= \nu_c / \sigma_l$ gives a measure of how narrow, or `coherent' the Lorentzian is.  Here, instead of subtracting the Poisson (`white') noise in each PSD band we model this with an additive constant, $C$.  The hard and soft band have similar shape, so we jointly model the data, allowing certain parameters untied in model variants described below.  The models are fit by minimising the $\Large \chi^2$ statistic within the package \textsc{xspec}.  The parameter and $\Large \chi^2$ space were explored with \citet{goodmanweare2010} Markov Chain Monte Carlo (MCMC) sampling.  Quoted parameter errors are from the $90 \%$ credible intervals.  We used a linear binning factor of $5$ to improve the frequency resolution in the modelling.  With a time bin $dt = 50$\,s this results in a combined total of $1602$ data bins for the soft and hard PSDs.  The resulting PSDs and model fits along with the model standard residuals are shown in Figure~\ref{fig:psdmod}.

The Akaike information criterion was used for model selection to compare models with additional complexity (\citealt{akaike74}).  The corrected form of this statistic takes account for the bias from the small sample size, which can be expressed as

\begin{equation}
\label{eqn:aic}
            {\rm AICc} = 2k + {\chi}^2 + \frac{2k^2 + 2k}{n - k - 1}
\end{equation}

\noindent where $k$ is the number of free model parameters, $n$ is the number of data bins in the PSD and $\chi^2$ is used to represent the negative of twice the logarithm of the likelihood of the model.  The model with the lowest ${\rm AICc}$ is the most preferred over all models.  The ${\rm \Delta AICc} = {\rm AICc} - \min({\rm AICc})$, with a ${\rm \Delta AICc} < 2 $ suggesting weak evidence in favour of the model. A ${\rm \Delta AICc} \sim 3-7$ suggests moderate evidence for the model and $>10$ implies the model is strongly favoured (e.g. \citealt{burnhamanderson2007}).  

We initially model the PSDs using a single broad Lorentzian component (model: Broadband), with all parameters free, giving $\Large \chi^2$/dof (degrees of freedom) $=1884.9/1594$.   The $\Large \chi^2$ and AICc are given in Table~\ref{tab1} and the residuals of this model are shown in Figure~\ref{fig:psdmod}.  The next more complex model has the addition of a narrower Lorentzian ($l_2$) at $\sim 2 \times 10^{-4}$\,Hz (model: BB+1Lor).  With ${\nu}_{\rm c}$ and $\sigma_{\rm l}$ tied between the two PSDs and $N_{\rm hard}$ and $N_{\rm soft}$ free, this improves the fit by $\Delta \Large \chi^2 = 22.7$ for 4 additional dof.  Compared to the `Broadband' model the ${\rm \Delta AICc} = 14.6$, indicating a strong preference for the more complex model.  Allowing any of the parameters to be untied between the two PSDs does not notably improve the model fit. 

The addition of a Lorentzian ($l_3$) with ${\nu}_{\rm c} \sim 8 \times 10^{-4}$\,Hz (model BB+2Lor) reduces the $\Large \chi^2 = 14.9$ for 4 additional dof.  Untying the parameters of the additional Lorentzian between the two PSDs does not significantly improve the fit.  Compared to the previous model the ${\rm \Delta AICc} = 10.9$, again suggesting the more complex model is strongly supported.

The most complex model we test involves the addition of a narrow Lorentzian ($l_1$) with ${\nu}_{\rm c} \sim 1 \times 10^{-4}$\,Hz (model: BB+3Lor).  With all but the normalisation parameters tied across the two energy bands, the best fit model gives $\Large \chi^2$/dof $=1824.1/1584$, with 4 additional parameters compared to the previous model.  The ${\rm \Delta AICc} = 14.98$, again suggesting the more complex model is strongly supported.  The low sampling resolution at low frequencies means the fit is largely insensitive to ${\nu}_{\rm c}$ of the broadband Lorentzian, so we fix this to zero (model: BB+3Lor($\nu_{\rm c_{\rm BB}}=0$)). This gives no notable change to the fit, but helps remove some parameter degeneracies between ${\nu}_{\rm c_{\rm BB}}$ and $\sigma_{\rm BB}$.  Equally, ${\nu}_{\rm c,l_1}$ and $\sigma_{\rm l_1}$ (i.e. the Lorentzian at ${\nu}_{\rm c} \sim 1 \times 10^{-4}$\,Hz also display degeneracies, so we fix these at approximately the median MCMC values of $Q= \nu_c / \sigma_l = 5$ (model: BB+3Lor($Q_{\rm l_1}=5$)).  The parameters for this model are shown in Table~\ref{tab2}.  The addition of more Lorentzian components provides an improvement of ${\rm \Delta AICc} < 2$, so we do not consider those models further. 

For the no-dip segments, a single zero-centered Lorentzian provides an adequate description of the hard band PSD.  The additional parameters are $\sigma_{\rm no-dip,hard} = 4.55${\raisebox{0.5ex}{\tiny$_{-2.28}^{+1.25}$}}\,$ \times 10^{-5}$\,Hz and $N_{\rm no-dip,hard} = 0.29${\raisebox{0.5ex}{\tiny$_{-0.07}^{+0.06}$}}.   Additional components do not significantly improve the AICc so we do not explore this further.

\begin{table}
\centering
\begin{tabular}{lcccc}
\hline
Model  & $N_{\rm pars}$ & $\Large \chi^2$ / d.o.f    & ${\rm AICc}$  & ${\rm \Delta AICc}$ \\
$(1)$ & $(2)$ & $(3)$ & $(4)$ & $(5)$ \\
\hline
Broadband (BB)              & 8         & 1884.9 / 1594 & 1900.9 & 40.5   \\
BB+1Lor                     & 12        & 1862.2 / 1590 & 1886.3 & 25.9   \\
BB+2Lor                     & 14        & 1847.2 / 1588 & 1875.5 & 15.1   \\
BB+3Lor                     & 18        & 1824.1 / 1584 & 1860.5 & 0.1    \\
BB+3Lor ($\nu_{\rm c_{\rm BB}}=0$) & 17        & 1826.4 / 1585 & 1860.8 & 0.4    \\
BB+3Lor ($Q_{\rm l_1}=5$)   & 16        & 1828.0 / 1586 & 1860.4 & *min*  \\
\hline
\end{tabular}
\caption{Fit statistics for the PSD model to the hard and soft band dip segments.  Columns (1) states the model, (2) the number of free parameters, (3) is the $\Large \chi^2$ / d.o.f, (4) is the corrected Akaike criterion, ${\rm AICc}$, and (5) is the ${\rm \Delta AICc}$ compared to the minimum model.}
\label{tab1}
\end{table}

\begin{table}
\centering
\begin{tabular}{lccccc}
\hline
Comp. & $\nu_{\rm c}$ & $\sigma_{\rm l}$ & $N_{\rm hard}$ & $N_{\rm soft}$ & Q \\
 & $\times 10^{-4}$ & $\times 10^{-4}$ & $\times 10^{-1}$ & $\times 10^{-2}$ & - \\
 $(1)$ & $(2)$ & $(3)$ & $(4)$ & $(5)$ & $(6)$ \\
  \hline
$l_{\rm BB}$ & $0$ fixed                                         & $0.37${\raisebox{0.5ex}{\tiny$_{-0.28}^{+0.25}$}} & $1.80${\raisebox{0.5ex}{\tiny$_{-0.70}^{+0.90}$}}   & $7.95${\raisebox{0.5ex}{\tiny$_{-2.89}^{+0.04}$}}  & - \\
$l_1$ & $0.89${\raisebox{0.5ex}{\tiny$_{-0.14}^{+0.04}$}}  & $0.18$                                            & $0.89${\raisebox{0.5ex}{\tiny$_{-0.47}^{+0.37}$}}  & $0.49${\raisebox{0.5ex}{\tiny$_{-0.28}^{+1.02}$}} & 5f  \\
$l_2$ & $2.16${\raisebox{0.5ex}{\tiny$_{-0.23}^{+0.08}$}} & $1.20${\raisebox{0.5ex}{\tiny$_{-0.54}^{+0.84}$}} & $1.05${\raisebox{0.5ex}{\tiny$_{-0.46}^{+0.31}$}} & $2.44${\raisebox{0.5ex}{\tiny$_{-1.06}^{+0.01}$}}  & 2 \\
$l_3$ & $6.95${\raisebox{0.5ex}{\tiny$_{-6.01}^{+1.43}$}} & $7.07${\raisebox{0.5ex}{\tiny$_{-2.0}^{+4.2}$}}   & $0.28${\raisebox{0.5ex}{\tiny$_{-.12}^{+.53}$}}  & $2.0 ${\raisebox{0.5ex}{\tiny$_{-1.1}^{+1.0}$}}  & 1 \\
\hline
\end{tabular}
\caption{Table for the best fitting model parameters for the BB+3Lor ($Q_{\rm l_3}=5$) model.  Errors are from the MCMC posterior $90 \%$ credible intervals.  This model consists of a broad zero-centered Lorentzian (BB) plus 3 narrower Lorentzians ($l_1 , l_2, l_3$) plus an additive constant $P_{\rm n}$ for Poisson noise.  All components are tied across energy bands except $N_l$ and the noise levels, $P_{\rm n, hard} = 23.95${\raisebox{0.5ex}{\tiny$_{-0.45}^{+0.34}$}}, and $P_{\rm n, soft} = 9.53${\raisebox{0.5ex}{\tiny$_{-0.16}^{+0.14}$}}}
\label{tab2}
\end{table}


\subsection{Periodogram of dipping observations}

\begin{figure}
          \vspace*{-8pt}
          \hspace*{-16pt}
          \includegraphics[width=0.52\textwidth,angle=0]{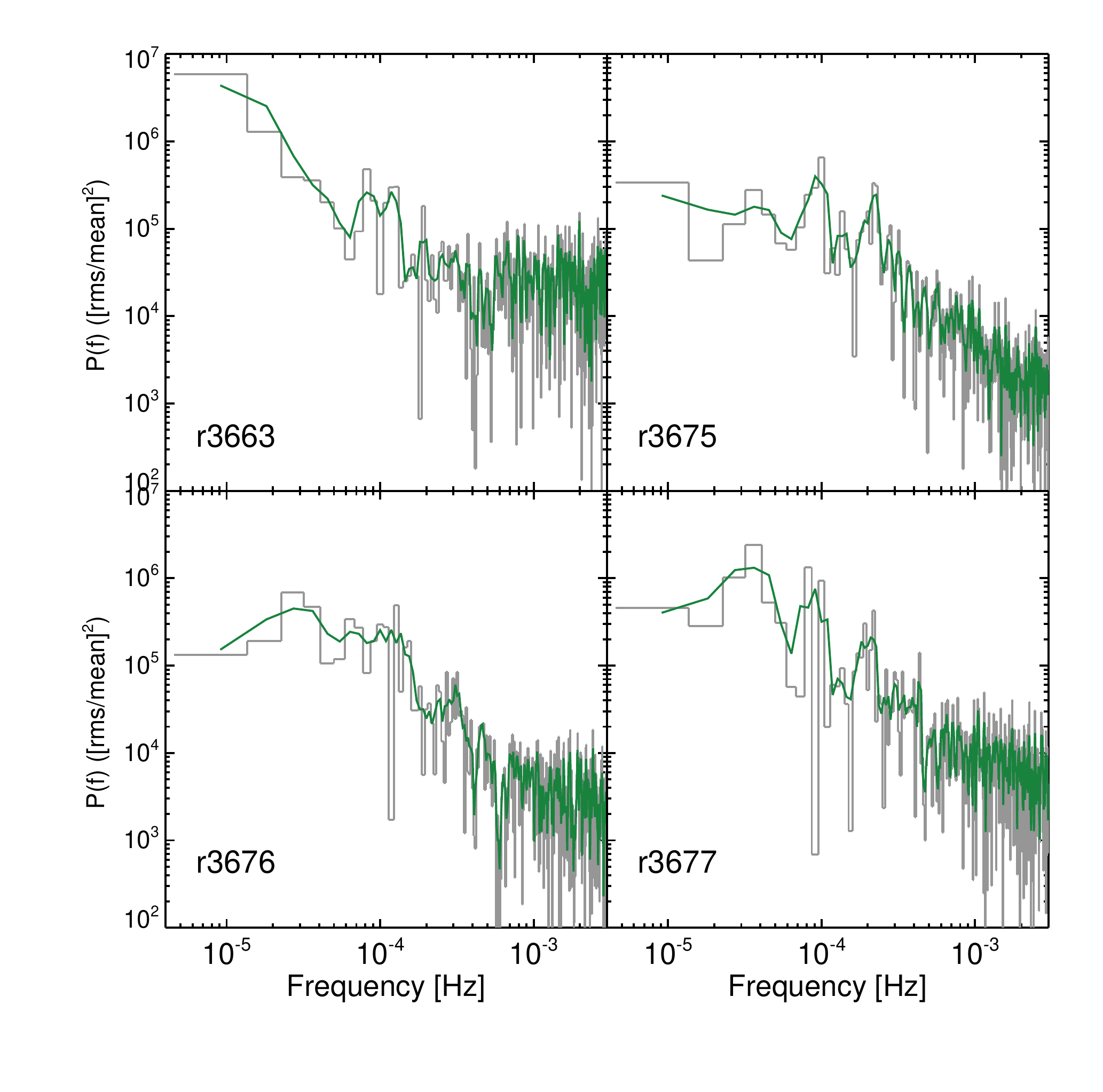}
          \vspace*{-25pt}
      \caption{Periodograms for the hard band dipping observations (without Poisson noise subtraction).  The green curve is the Gaussian smoothed periodogram over $n=3$ neighbouring bins.}
       \label{fig:per}
   \end{figure}

To check that no one particular observation is driving the peaks in the average PSD we estimate the unbinned periodogram.  Figure~\ref{fig:per} shows the periodogram for the four dipping observations.  It is well known that the periodogram estimate at each frequency follows a ${\Large \chi}^2_{\nu=2}$ distribution so large scatter around the true underlying power spectrum is expected (e.g. \citealt{priestley81}).  We therefore also show a Gaussian smoothed with periodogram over $n=3$ neighbouring frequency bins.  Peaks in power at $\sim 1 \times 10^{-4}$\,Hz and $\sim 2 \times 10^{-4}$\,Hz are found in each periodogram.  Another peak or bend can be seen at $\sim 0.4 \times 10^{-4}$\,Hz in all but rev~$3663$, which shows power at lower frequencies.

We refit these periodograms using the maximum likelihood methods (MLE) for unbinned periodograms outlined in \citet{vaughan10} and \citet{BarretVaughan12}.  The power spectral modelling results are consistent with those found in Section~\ref{sec:psdmod}, so we do not show those here.  For rev~$3676$ the Lorenztzian centroids  $l_1$ and $l_2$ appear to shift to higher frequencies, however modelling reveals these to be consistent within error bars with the other revs.  The only notable difference is the increase in power at low frequencies in rev~$3663$, which requires a zero-centered Lorentzian with width $\sigma_{l_{\rm BB}} \sim 1 \times 10^{-5}$\,Hz.  Removing rev~$3663$ from the combined PSD modelling in Section~\ref{sec:psdmod} does not significantly alter the results, so we choose to leave it in to increase the number of estimates in each frequency bin.

\subsection{Change point analysis}

In order to study what aspect of the dips is causing the peaks in the power spectra, we modelled the abrupt light curve changes in the dip segments using change point analysis.  The \texttt{R} package \texttt{bcp}\footnote{\url{https://cran.r-project.org/web/packages/bcp/}} was used which implements the Bayesian change point method of \citealt{wangemerson2015bcp}.  This partitions the light curve into constant mean `blocks' between change points.  MCMC sampling is used to model the posterior probability at each change point, $p_{\rm ch}$, defined between [0,1].  Given the sharp changes in the light curves when dipping occurs, this method is particularly well suited to the data.

The Poisson noise starts to dominate at $\sim 2 \times 10^{-3}$\,Hz so a time bin of $dt=500$\,s was used in order to smooth over any noise variations.  A posterior probability $p_{\rm ch} >0.4$ was used to identify significant change point locations.  From these we determine the dip onset time $t_{\rm 0}$, dip duration $t_{\rm d}$, and dip midpoint, $t_m = (t_{\rm 0}+t_{\rm d})/2$.  The period $P$ is then taken as the time between successive dip midpoints.  

\begin{figure}
    \centering
            \hspace*{0pt}
            \includegraphics[width=0.47\textwidth,angle=0]{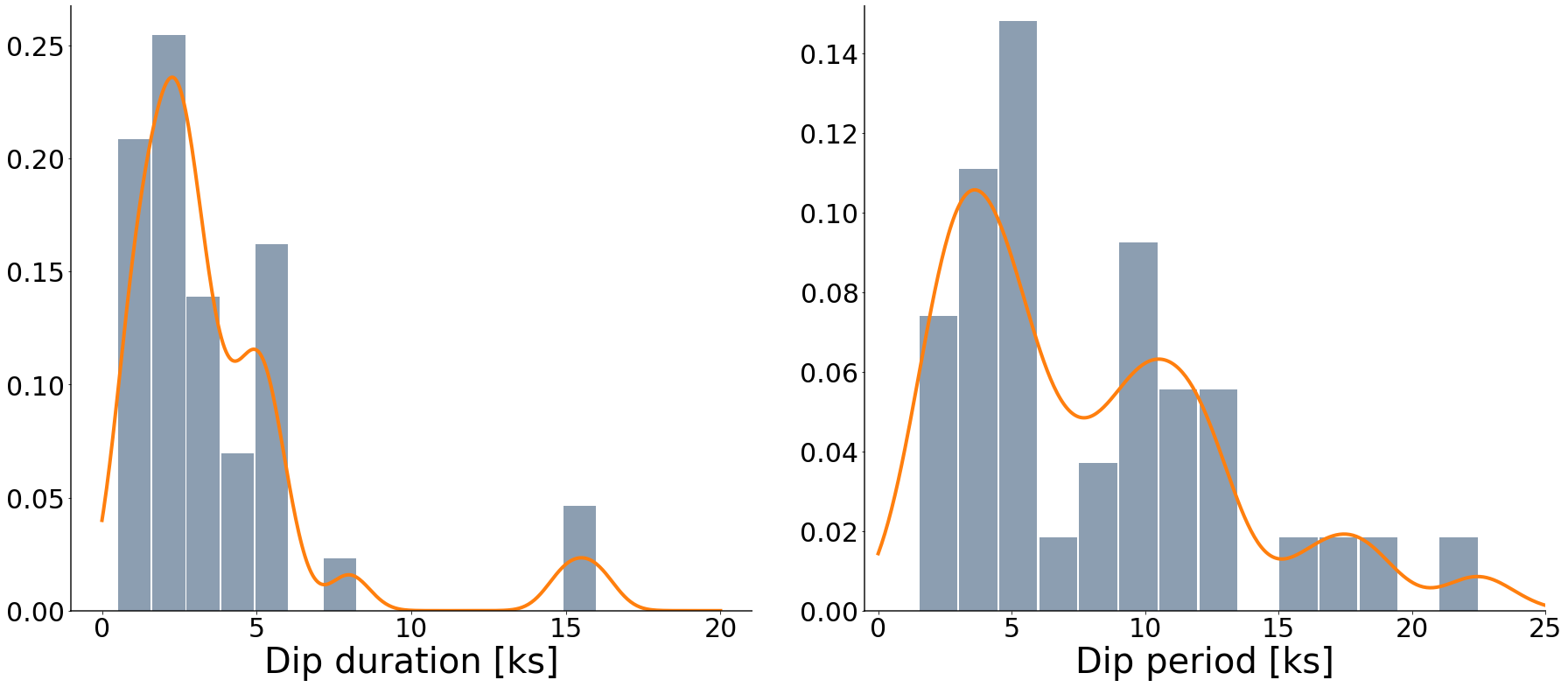}
            \vspace*{-0pt}
        \caption{Density of dip duration, $t_{\rm d}$, (left) and dip period, $P$, (right) as estimated using Bayesian change point analysis.  The Gaussian kernel density estimate is shown in orange.}
        \label{fig:dist}
    \end{figure}

Figure~\ref{fig:dist} shows the distribution of $t_{\rm d}$ and $P$ for all four dipping observations.  A Gaussian kernel density estimate is overlaid on the observed distributions.  The distribution of $t_{\rm d}$ peaks at $\sim 3$\,ks and falls off approximately log-normally with increasing dip duration, i.e. there is a much larger probability for shorter dips.  The distribution of $P$ is multi-modal with peaks at $\sim 5$ and $10$\,ks, with a smaller probability of periods at $\sim 17$\,ks.  A similar distribution is observed if we represent the $P$ as the time between successive $t_{\rm 0}$, suggesting no relationship between the duration of the dips and the dip period.


\subsection{Variability spectra}
\label{sec:varspec}

We study the energy dependence of the variability at a given timescale using frequency resolved covariance spectra (\citealt{WilkinsonUttley09}).  Using a high S/N reference band, the correlated variability is picked out in a given comparison band, thus improving the S/N of the variability spectrum compared to traditional rms spectra.  We compute the covariance spectra in the Fourier domain (\citealt{uttley2011}; \citealt{CassatellaETAL2012}; \citealt{uttley14rev}).  The $0.3-4.0$\,keV band is used as the broad reference band, but the choice of this band does not affect the shape of the resultant covariance spectra.

\begin{figure}
\centering
\hspace*{-10pt}
\includegraphics[width=0.5\textwidth,angle=0]{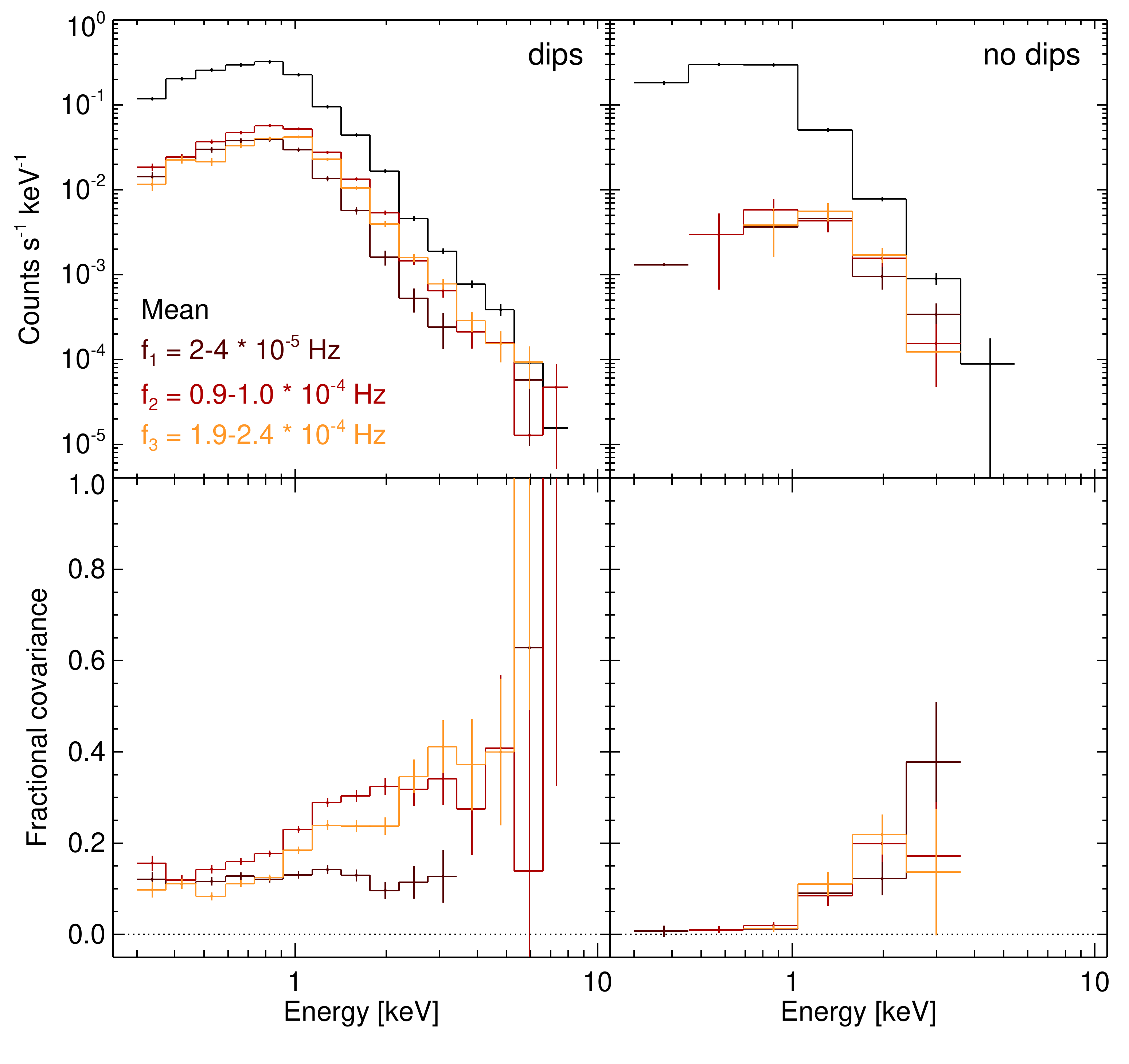}
\vspace*{-12pt}
    \caption{Variability spectra for the dip (left) and no-dip (right) segments. The frequency ranges correspond to the low frequency, the first and the second dip peaks respectively.  The upper panels show the time averaged and variability spectra in absolute units.  The lower panels show the fractional covariance spectra for each frequency band.  Missing bins are due to high Poisson noise in a given energy band.}
\label{fig:varspec}
\end{figure}

We estimate covariance spectra in three frequency bands: $f_1 = 2.0-4.0 \times 10^{-5}$\,Hz, $f_2 = 0.9 -1.1 \times 10^{-4}$\,Hz and $f_3 = 1.9 - 2.4 \times 10^{-4}$\,Hz.  These timescales probe the long timescale spectral variability as well as the spectral variability associated with the two prominent dipping timescales.  Figure~\ref{fig:varspec} shows the resulting variability spectra.  The top panels shows these data in absolute units (${\rm cts}~{\rm s}^{-1}~{\rm keV}^{-1}$) whilst the lower panels show the variability spectra in fractional ($\%$) units.  In the dipping segments, the long timescale spectral variability is approximately flat, with a $10 \%$ rms amplitude.  The dipping timescales, $f_2$ and $f_3$, show an approximately linear dependence with energy, with larger variability amplitude at higher energies.  This is true of all frequency bands up to higher frequencies, but we do not show those here.

For the dipping segments, as an additional check, we calculate the variability amplitude of each of the components in the preferred model (BB+3Lor $Q_{\rm l_3}=5$) by integrating the rms in each Lorentzian component.  This shows the same trend in energy space as the frequency dependent covariance spectra.  For the no-dip segments, the fractional variability amplitude is almost below $1$\,keV, with a sharp increase in rms for energies above this.

\section{Discussion}
\label{sec:disco}


\subsection{The dipping behaviour}

The light curves of NGC 247 ULX-1 display deep and quasi-periodic dips imprinted on top of stochastic noise variations.  Modelling the PSD with multiple Lorentzians reveals three preferred timescales for the dipping behavior.  The peak in power of a Lorentzian occurs at frequency $\nu_{\rm max} = \sqrt{\nu_c^2 + \sigma_l^2}$ (see e.g. \citealt{belloni2002lormax}).  For the preferred model, this gives $\nu_{\rm max,BB} \sim 0.4 \times 10^{-4}$\,Hz, $\nu_{\rm max,l_1} \sim 0.9 \times 10^{-4}$\,Hz, $\nu_{\rm max,l_2} \sim 2 \times 10^{-4}$\,Hz and $\nu_{\rm max,l_3} \sim 9 \times 10^{-4}$\,Hz.  The coherence between the hard and soft bands is high ($\gsim 0.9$) at the Lorentzian peaks and drops where the Lorentzians overlap, providing further evidence for multiple dipping timescales. 

Time series change point analysis reveals the dip periods, $P$, cluster around $\sim 5$ and $10$\,ks.  This distribution of $P$ is consistent with the peak timescales observed in the power spectra (see Figure~\ref{fig:psdmod}).  The Lorentzian $Q$ values tell us that there is some scatter around a preferred periodicity, as is observed in the change point period analysis.  The dip durations $t_d$ preferentially take on values between $1-5$\,ks, which will add some additional broadening to the measured Lorentzian $Q$ values.  There are several broad dips with $t_d > 15$\,ks, which clearly stand out from the other measured dip durations in Figure~\ref{fig:dist}.  This could be suggesting an alternative mechanism or timescale for these longer duration dips.

The change point analysis does not find any periods with $P < 1000$\,s suggesting this analysis is either insensitive to this dips, e.g. they could be of lower amplitude and therefore harder to detect, in agreement with the lower amplitude of the Lorentzian $l_3$.  Alternatively, this high frequency component is related to an increase in the broadband noise continuum, which appears when the source enters the higher-flux / dipping epochs.

The time averaged and frequency resolved covariance spectra in Figure~\ref{fig:varspec} show that the variations are larger at higher energies in both the dip and no-dip epochs.  The exception being the $f_1 = 2.0-4.0 \times 10^{-5}$\,Hz frequency band in the dip epochs.  Here, the fractional covariance is independent of energy.  The coherence between the soft and hard bands (see Figure~\ref{fig:psdcoh}) also drops to $\sim 0.6$ at these frequencies.  This suggests that the long timescale variability is different to the dip generating mechanism at shorter timescales.


\subsection{Origin of the dips in NGC 247 ULX-1}

Assuming a black hole with $M_{\rm BH} = 10 M_{\odot}$, the peak timescales $\nu_{\rm max} \sim 0.5 - 9 \times 10^{-4}$\,Hz correspond to an orbital radius ($^{3}\sqrt{GMt^2/4\pi^2}$) of $10^{11}$\,cm, or $2-10 \times 10^4$\,$R_G$ (where $R_G = GM/c^2$).  A similar value is obtained for the case of a NS compact object.  In both cases, this region is very distant from the compact object, suggesting an origin in the outer accretion flow.  Two likely scenarios for the dips are warping of an accretion disc seen close to edge on, periodically obscuring the central source each orbit, or, the launching of a wind following changes in local $\dot{M}$.  Alternatively, a combination of both processes could be causing the dips at different timescales.

The radius associated with Compton scattering through a medium around the spherisation radius can be expressed as $R_{\rm C} = \sqrt{c t_{\rm esc} / \sigma_{\rm T} n_{\rm e}}$ (where $c$ is the speed of light, $\sigma_{\rm T}$ is the Thomson scattering cross-section, $n_{\rm e}$ is the electron density, and $t_{\rm esc}$ is the timescale for photons to escape) would be of the order of $10^9$\,cm, assuming a disc-like $n_{\rm e}$ density ($10^{20}$\,cm$^{-3}$), while such radius would be comparable ($10^{11-12}$ cm) if we adopt a wind-like density ($10^{14}$ cm$^{-3}$).  On the other hand, the wind radius, $R_{\rm wind}=L_{\rm ion}/n_{\rm e}\xi$ (where $L_{\rm ion}$ is the ionising luminosity and $\xi$ the ionisation parameter), that we estimate from the best-fit parameters of the \xmmn RGS-EPIC joint fit is also around $10^{11-13}$\,cm (see \citealt{pinto2021_ngc247sub}).  The lower density assumption is favoured as this agrees with both the outflow and dip timescale properties.

The mid plane of the accretion disc, i.e. the spherisation radius where the disc height is comparable to the distance from the accretor, is instead $R_{\rm sph}= 5/3 \dot{M} R_{\rm in} \sim 100$\,$R_G$ (assuming a standard $R_{in} = 6$\,$R_{G}$ and $\dot{M} \sim 10$\,$\dot{M}_{\rm Edd}$, which would be sufficient to achieve $L_X \sim 5 \times 10^{39} \ergs$ for $M_{\rm BH} = 10 M_{\odot}$).  This means that the timescales of the variability detected in NGC 247 ULX-1 are likely associated with the outer disc and, most likely, with the region where the wind becomes optically thin (e.g. $n_{\rm e} \lesssim 10^{14}$\,cm$^{-3}$) and rather than with the inner accretion flow or the optically thick wind ($10^{20}$\,cm$^{-3}$) which would be expected near the spherisation radius.  For a comparison, the wind photosphere in the Galactic super-Eddington source SS\,433 is $\sim 10^{12}$\,cm (see, e.g. \citealt{Fabrika2004}), which further indicates that NGC 247 ULX-1 quasi-periodic dips might be associated with some structure present in the outer disc where the photosphere becomes optically thin.

The presence of dips and their long timescales argue in favour of a moderate-to-high inclination in this source, which agrees with the scenario in which soft ULXs are primarily seen along the edge of the wind.  The dips seen in both BH and NS X-ray Binaries (XRBs) are believed to be caused by periodic obscuration of the central X-ray source by structure located in the outer regions of a disc.  These dips are observed in sources at high inclination ($i>60$\,deg, e.g. \citealt{franketal1987}).  For NGC 247 ULX-1, \citet{yaofeng2019ngc247} derived an $i >67$\,deg via modeling the UV/optical spectral energy distribution (SED) with an irradiation model, providing further evidence for a high inclination viewing angle in this source.  Additionally, winds observed in BH or NS X-ray binaries (XRBs) have been mostly  observed in sources in which the disc is inclined at a large angle to the line of sight (e.g. \citealt{ponti2012winds, diaztrigo2016}). 

Given the difficulty to access the inner regions of the accretion flow, mostly obscured by the scale height of the disc at the spherisation radius and the wind, we are not surprised that we did not detect the time lags between the soft and hard band that have been found in other, harder, ULXs. At these angles, the photons would undergo loads of Compton scattering, losing the original information.

If the dips are caused by a wind structure, what is not clear is whether the short-term variations in optical depth are just stochastic variability (at a given $\dot{M}$) due to the clumpy nature of the wind.  Alternatively, each variation corresponds to a change in the underlying $\dot{M}$ variability in the local accretion rate where the wind is launched.  This variability in the wind launching would thereby cause these quasi periodic dips, rather than being pre-existing structure with the appropriate scale height.  This idea is supported by the onset of dipping behaviour in the higher flux observations.  In fact, the bolometric luminosity of SSUL sources slightly increases when the blackbody temperature decreases and the blackbody radius increase. This behaviour is common to several other SSUL (e.g. \citealt{Feng2016}; \citealt{Urquhart2016}).

The broadband spectral model decomposition in D'A\`{i} et al, \textit{ in prep} reveals two broad thermal components.  These are interpreted as emission from an extended, optically thick, cold photosphere, and a hotter emission component, originating from regions closer to the compact object.  In the `out of dip' epochs (i.e. when a dip is not occurring), the short term variations are predominantly driven by changes in the flux of the hotter component. During the dips, both broad spectral components show significant decrease in their flux, particularly the hotter component.  One scenario, where the dipping mechanism is caused by occultation further out than the typical radii where both components originate, would be consistent with what we infer from the timing properties in the present paper.

An alternative explanation for the dips is that of intrinsic luminosity variations due to the propeller effect switching on-and-off accretion in a magnetised NS (see e.g. \citealt{Revnivtsev2015rev} for a review).  This was suggested as the origin of large amplitude variations in the X-ray light curve of the ULX M\,82 X-1 by \citet{Tsygankov2016}, however these occur on timescales of $\gsim$\,weeks.  Interestingly, the `hiccuping' X-ray light curve of the accreting millisecond X-ray pulsar, IGR J18245-2452, displays a remarkable resemblance to those in NGC 247 ULX-1 (\citealt{ferrigno2014hiccup}), However, the dips in this source show complex spectral hardening, whereas in NGC 247 ULX-1 the dips are spectrally harder (see \citealt{pinto2021_ngc247sub} Figure~2).  The detection of pulsations in NGC 247 ULX-1 would provide more evidence for this scenario, which is scope for future work.

\subsection{Dips in other ULXs}

For NGC 55, \citet{stobbart2004} report several dips in the \xmm light curve.  These are fewer in number, but have a similar dip duration and separation between consecutive dips.  The authors argued the dips were likely due to material at $\gsim 2 \times 10^{11}$\,cm from the central object (assumed to be a BH).  At this radius the orbital timescale is $\approx 10^{4}$\,s, consistent with our PSD peak timescales.  This is in agreement with the idea that a disc warp or extended wind structure is causing the dips each orbit in NGC 247 ULX-1.

The two broad dips with $t_d \gsim 15$\,ks in observations rev~$3663$ and rev~$3677$ are similar to the broad deep dips seen in M 51 ULX-1 (\citealt{Urquhart2016eclipse}).  These authors argued that these dips are caused by eclipsing events and placed limits on the orbital period.  CG X-1, a ULX in the galaxy Circinus, has also shown multiple eclipsing events (\citealt{qiu2019eclipse}), with the authors finding an orbital period of $\sim 7$\,hrs.  It is possible that we are seeing a similar eclipses in NGC 247 ULX-1, meaning the separation of $\sim 19$\,days for these deep dips may correspond to some integer multiple of the orbital period.  If this is the case, then the dipping activity observed in NGC 247 ULX-1 may be triggered by the companion at some point in the orbital phase.  This remains scope for future work.


A tentative $\sim 2$\,hr ($\sim 1 \times 10^{-4}$\,Hz) quasi-periodic oscillation (QPO) was claimed in NGC 628 (\citealt{liu2005dips}).  The light curve of this source displays a similar dipping process to NGC 247 ULX-1.  The authors scaled this frequency to QPOs typically seen in black hole X-ray binaries (XRBs; see e.g. \citealt{remmc06}) and inferred a black hole mass $M_{\rm BH} \sim 10^3 M_{\odot}$.  However, the light curve properties of the ULX dips are strikingly different to QPOs seen in XRBs and active galaxies (e.g. \citealt{ingrammotta2020qporev}, \citealt{alston14b,alston15,alston16qpo}).  If the dips are caused by a similar process to NGC 247, then this would argue in favour of a stellar mass compact object in NGC 628.

\section{Conclusions}
A long \xmm observation of \ngc has revealed dipping in the light curves on $\sim 1-10$\,ks timescales.  This is consistent with the idea that supersoft ULXs are viewed from high inclination, with the dipping caused by an obscuration process from the outer accretion region.  These findings will motivate future studies on the source geometry and emission mechanisms at super-Eddington accretion.

\section*{Acknowledgments}

WNA is supported by an ESA research fellowship.  AD, MDS and FP acknowledge financial contribution from the agreement ASI-INAF n.2017-14-H.0 and INAF main-stream.  HPE acknowledges support under NASA contract NNG08FD60C.  TPR gratefully acknowledges support from the Science and Technology Facilities Council (STFC) as part of the consolidated grant award ST/000244/1.  This paper is based on observations obtained with \xmmns, an ESA science mission with instruments and contributions directly funded by ESA Member States and the USA (NASA). 

\section*{Data Availability}
All of the data underlying this article are publicly available from ESA’s \xmm Science Archive (XSA; \url{https://www.cosmos.esa.int/web/xmm-newton/xsa}) and NASA’s HEASARC archive (\url{https://heasarc.gsfc.nasa.gov/}).  Light curves are available from the author upon request.




\bibliographystyle{mnras}
\bibliography{var,bib_ulx} 



%
%
%
%
%


\bsp	
\label{lastpage}
\end{document}